# Band structure and carrier effective masses of boron arsenide: effects of quasiparticle and spin-orbit coupling corrections

Kyle Bushick, Kelsey Mengle, Nocona Sanders, and Emmanouil Kioupakis[a]

*Department of Materials Science and Engineering, University of Michigan, Ann Arbor, Michigan, 48109, USA*

We determine the fundamental electronic and optical properties of the high-thermal-conductivity III-V semiconductor boron arsenide (BAs) using density functional and many body perturbation theory including quasiparticle and spin-orbit coupling corrections. We find that the fundamental band gap is indirect with a value of 2.049 eV, while the minimum direct gap has a value of 4.135 eV. We calculate the carrier effective masses and report smaller values for the holes than the electrons, indicating higher hole mobility and easier p-type doping. The small difference between the static and high frequency dielectric constants indicates that BAs is only weakly ionic. We also observe that the imaginary part of the dielectric function exhibits a strong absorption peak, which corresponds to parallel bands in the band structure. Our estimated exciton binding energy of 43 meV indicates that excitons are relatively stable against thermal dissociation at room temperature. Our work provides theoretical insights on the fundamental electronic properties of BAs to guide experimental characterization and device applications.

   III-V semiconductors such as GaAs have underpinned a variety of technological advances such as transistors, solar cells, and solid-state lighting. Boron arsenide (BAs) is a largely unexplored III-V semiconductor that has recently attracted attention after theoretical predictions[1] of its ultrahigh thermal conductivity (second only to diamond as a bulk material) were experimentally confirmed.[2–4] While the basic structure of BAs has been studied since the 1960's,[5,6] the high quality of the synthesized samples in the latest studies promise to generate significant interest in the material for applications ranging from power electronics to optoelectronic devices. However, to determine the scope of possible applications of BAs, it is first necessary to determine its fundamental electronic and optical properties.

   Modern first-principles computational methods based on density functional theory (DFT) provide a predictive tool to theoretically investigate the properties of materials such as BAs that complements experiment. Indeed, the electronic band structure and related properties of BAs have been previously explored with DFT calculations.[7–11] However, the variety of computational methods employed and a dearth of experimental evidence – primarily a result of the difficulty to synthesize high quality BAs samples[3,7] – has led to inconclusive findings. Furthermore, few theoretical works have applied quasiparticle corrections or considered spin-orbit coupling effects in BAs. The two such works that include quasiparticle corrections limit their analysis to the band structure itself,[9,10] leaving numerous other important electronic properties of BAs such as absolute band positions with respect to vacuum, carrier effective masses, exciton properties, and dielectric constants largely unexplored.

   In this work, we conduct a detailed theoretical characterization of the fundamental electronic and optical properties of BAs using predictive atomistic calculations based on density functional and many-body perturbation theory. We calculate the electronic band structure including quasiparticle and spin-orbit coupling corrections to determine the location and magnitude of the direct and indirect band gaps. We further evaluate the carrier effective masses, exciton binding energy, the zone-center phonon frequencies, the static and high-frequency dielectric constants, and the imaginary part of the dielectric function due to direct optical absorption. Our results inform the experimental synthesis and characterization of this promising high-thermal-conductivity III-V semiconductor and guide the development of electronic and optoelectronic devices.

---

[a] Electronic mail: kioup@umich.edu

Our DFT calculations are based on the local-density approximation (LDA) as implemented in the Quantum ESPRESSO code.[12] We employed norm-conserving pseudopotentials for the valence electrons of boron and arsenic, a plane-wave energy cutoff of 70 Ry, and an 8×8×8 Monkhorst-Pack Brillouin-zone (BZ) sampling grid, which converged the total energy to within 1 mRy/atom. All calculations were performed for the zincblende structure using the experimentally reported lattice parameter of 4.7776 Å to avoid introducing errors by the DFT structural relaxation.[5] We subsequently utilized the $G_0W_0$ method (hereafter denoted GW) and the BerkeleyGW code to calculate quasiparticle corrections for the lowest eighteen bands of the electronic band structure.[13] On the GW level, the following parameters were used: an 8×8×8 BZ sampling mesh, a dielectric-matrix cutoff value of 30.0 Ry, and a sum over 700 and 717 bands for the dielectric-function and self-energy sums over unoccupied states, respectively. Using these settings, the GW absolute values of the quasiparticle energies were converged to within 5 meV. For both the LDA and GW calculations, we used the maximally localized Wannier function method[14] as implemented in the wannier90 package[15] to obtain finer band structures, including quasiparticle and spin-orbit coupling corrections as in Ref. 16. Eight Wannier functions were extracted out of these eighteen bands by projecting on bond-centered bonding and antibonding orbitals. The frozen energy window ranges from approximately 28 eV below the valence band maximum (VBM) to 4.5 eV above the conduction band minimum (CBM). We further applied spin-orbit coupling corrections by evaluating the spin-orbit Hamiltonian matrix elements and diagonalizing the resulting matrix.[17] From the interpolated band structures, we calculated effective masses for both the valence and conduction bands along various crystallographic directions. The effective-mass values were determined by fitting the band edge curves to the hyperbolic equation:

$$E(k) = \frac{\mp 1 \pm \sqrt{1 + 4\alpha \frac{\hbar^2 k^2}{2m^*}}}{2\alpha} + E_0, \quad (1)$$

where $E(k)$ is the energy of the band at reciprocal-space point $k$, $\alpha$ is a fitting parameter that quantifies the non-parabolicity of the band, $m^*$ is the effective mass, and $E_0$ is the energy at the band extrema ($E_{VBM}$ for the hole and $E_{CBM}$ for the electron). The upper (lower) sign corresponds to the CBM (VBM) fit. The imaginary part of the dielectric function with and without excitonic effects was calculated with the Bethe-Salpeter equation method within BerkeleyGW, using an 18×18×18 BZ sampling grid and a numerical broadening width of 0.2 eV. The exciton binding energy is approximated using the Bohr model. Density functional perturbation theory (DFPT) within Quantum ESPRESSO was used to determine the phonon frequencies at Γ for the Lyddane-Sachs-Teller relationship to obtain the static dielectric constant. To find the ionization energy and electron affinity, we aligned the bands with respect to the vacuum level using a slab composed of 8 atomic layers along the nonpolar [110] direction with a slab thickness of 11.32 Å. To prevent interactions between periodic layers, we included 30 Å of vacuum space normal to the slab. The plane-averaged electrostatic potentials (ionic plus Hartree terms) were calculated for both the slab and bulk BAs, allowing us to reference the GW+SO eigenvalues to the average bulk electrostatic potential, as well as determine the energy difference between the average electrostatic potential in the middle layers of the slab and the vacuum level.[18]

The full electronic band structure calculated with the LDA and GW methods, as well as with and without spin-orbit coupling effects, is shown in Fig. 1. We find that in all cases the fundamental band gap of BAs is indirect in nature. The valence band maximum (VBM) occurs at the Γ point, while the conduction band minimum (CBM) is found along the Γ-X direction at k-point Δ with crystal coordinates (0.39, 0.00, 0.39). We find that the band-gap value is 1.190 eV for LDA while the GW corrections yield a value of 2.115 eV. Including spin orbit effects decreases the LDA band gap to 1.124 eV and the GW band gap to 2.049 eV. We find the smallest direct gap to occur at Γ; the LDA gap is 3.328 eV, the LDA+SO gap is 3.128 eV, the GW gap is 4.336 eV, and the GW+SO gap is 4.135 eV. These values are presented in Table 1. The discrepancy between our calculated GW+SO gap and the experimental measurements may originate from temperature or zero-point motion effects,[19] or they could be caused by the challenges in the growth of BAs samples. Our calculated values for the ionization energy and electron affinity are 6.544 eV and 4.495 eV, respectively. However, we note that work by Shaltaf *et al.* has found that using the plasmon-pole model for GW quasiparticle corrections can lead to inaccuracies in these band alignment calculations.[20] Thus, future work should examine these values with full frequency GW calculations. Additional energies for special points in the first Brillouin zone can be found in Table 2. We find that the GW corrections to the LDA band structure rigidly increase the band gap, while the curvature of the bands near the edges remains approximately the same between the two calculations. Furthermore, as can be seen in Figs. 1a and 1c, the inclusion of spin-orbit effects does not affect the overall shape of the band structure; however as seen in Figs. 1b and 1d, spin-orbit effects do result in the top three degenerate valence bands splitting by about 0.2 eV at $\Gamma$. We also find that the conduction bands at $\Gamma_{7c}$ and $\Gamma_{8c}$ split, mirroring the behavior of the valence bands at $\Gamma$. Additional spin-orbit effects around the $\Gamma$ point include an avoided crossing in the conduction band, as well as changes to the valence band curvatures.

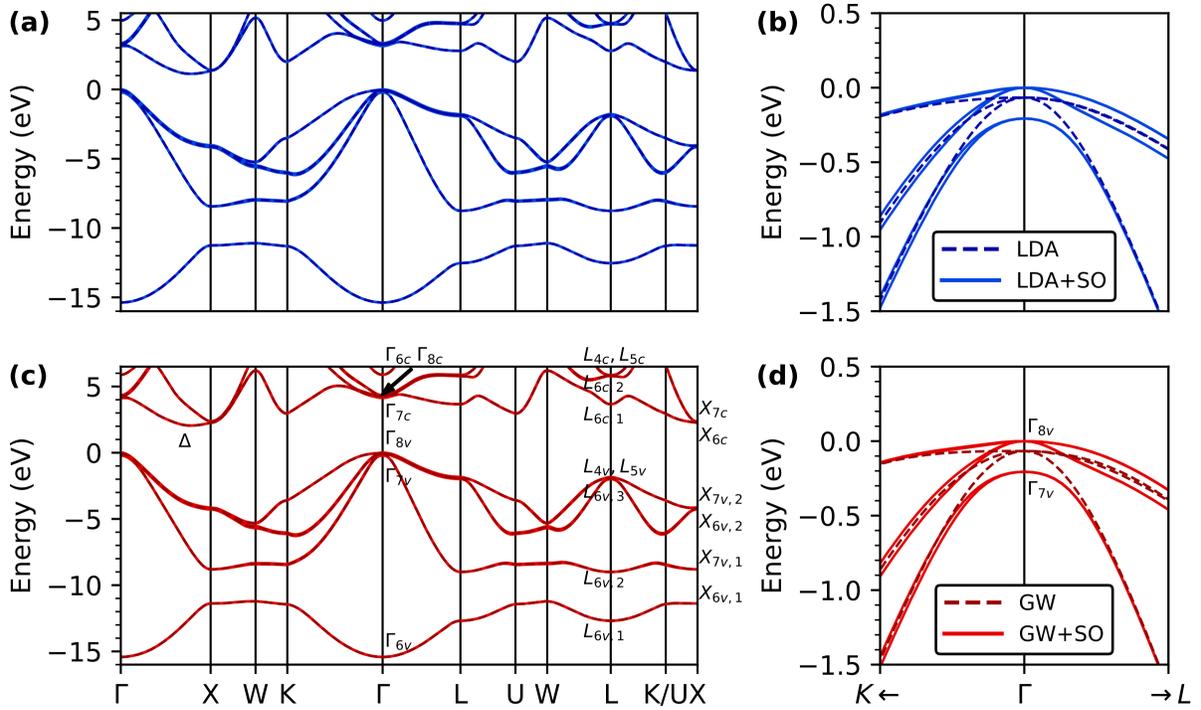

Figure 1: The band structure of BAs as calculated with the LDA (a-b) and the GW (c-d) method. The band structures are plotted both with (solid lines) and without (dashed lines) spin-orbit effects considered. Panels (b) and (d) focus on the spin-orbit splitting of the topmost valence bands. In each case the bands are referenced to the VBM location after the inclusion of spin-orbit effects.

Table 1: Our calculated values for the band gap of BAs (in eV) as obtained with different calculation methods compared to previous theoretical and experimental data from the literature.

| Calculation method | Indirect band gap | Minimum direct gap |
|---|---|---|
| This work, LDA | 1.189 | 3.328 |
| This work, LDA+SO | 1.124 | 3.128 |
| This work, GW | 2.115 | 4.336 |
| This work, GW+SO | 2.049 | 4.135 |
| Previous theory (GW) | 1.6[a], < 1.86[b] | 4.2[a], 4.0[b] |
| Previous experiment* | 0.67[c], 1.46[d] | 1.46[c] |

[a] Ref. 9
[b] Ref. 10
[c] Ref. 21, note this is an amorphous film
[d] Ref. 22
* Note that Ref. 6 also reports a value of 1.46 eV for the band-gap, but it is not clear if it is direct or indirect

Table 2: Calculated band energies (in eV) of special Brillouin-zone points (denoted in Fig. 1) of BAs. The LDA values are referenced to the LDA energy of the valence band at $\Gamma_{8v}$. The GW and GW+SO values are referenced to the GW+SO energy of the valence band at $\Gamma_{8v}$.

| Special Point | LDA | GW | GW+SO |
|---|---|---|---|
| $\Delta$ | 1.190 | 2.049 | 2.049 |
| $X_{6v,1}$ | -11.192 | -11.384 | -11.384 |
| $X_{7v,1}$ | -8.374 | -8.806 | -8.808 |
| $X_{6v,2}$ | -4.029 | -4.205 | -4.275 |
| $X_{7v,2}$ | -4.029 | -4.205 | -4.135 |
| $X_{6c}$ | 1.419 | 2.264 | 2.264 |
| $X_{7c}$ | 1.444 | 2.356 | 2.356 |
| $\Gamma_{6v}$ | -15.308 | -15.423 | -15.423 |
| $\Gamma_{7v}$ | 0.000 | -0.066 | -0.206 |
| $\Gamma_{8v}$ (x2) | 0.000 | -0.066 | 0.000 |
| $\Gamma_{7c}$ | 3.328 | 4.270 | 4.135 |
| $\Gamma_{8c}$ (x2) | 3.328 | 4.270 | 4.341 |
| $\Gamma_{6c}$ | 5.031 | 5.882 | 5.882 |
| $L_{6v,1}$ | -12.469 | -12.688 | -12.688 |
| $L_{6v,2}$ | -8.696 | -9.002 | -9.003 |
| $L_{6v,3}$ | -1.788 | -1.910 | -1.985 |
| $L_{4v}, L_{5v}$ | -1.788 | -1.910 | -1.837 |
| $L_{6c,1}$ | 2.841 | 3.658 | 3.658 |
| $L_{6c,2}$ | 4.811 | 5.826 | 5.789 |
| $L_{4c}, L_{5c}$ | 4.811 | 5.826 | 5.864 |

We subsequently evaluated the electron and hole effective-mass values using the GW+SO band structure data; a comprehensive list of effective masses is presented in Table 3 and is in agreement with work done by Nwigboji et al.[11] We find that along all directions the valence band splits into the heavy hole (hh), light hole (lh), and spin-orbit hole (soh). The spin-orbit splitting along the Γ-K direction results in six nondegenerate bands due to the lack of inversion symmetry in BAs. We find that the effective masses of holes are relatively small compared to the electron effective masses, which implies a lower acceptor ionization energy and thus easier p-type doping, as well as a potentially higher hole mobility.[23] The low hole effective masses in BAs compared to other III-V semiconductors can be physically understood by considering the smaller lattice constant, which increases the overlap between neighboring atomic orbitals, as well as the orbital character of the conduction and valence bands. The VBM is made up of 57.6% boron p-orbitals and 40.2% arsenic p-orbitals. This relatively covalent charge distribution contributes to large interatomic matrix elements between nearest-neighbor B and As atoms and thus a small mass. The CBM is made up of 30.9% boron p-orbitals, 24.0% boron s-orbitals, 20.1% arsenic p-orbitals, and 15.4% arsenic s-orbitals. The narrower spatial localization of the s-orbitals that form the CBM contributes to the heavier effective mass.

The relatively covalent nature of BAs is further revealed by its vibrational frequencies and dielectric constants. To obtain the phonon frequencies we relaxed the structure until the stress in each cell direction was less than ~3×10$^{-6}$ Ry/bohr$^3$, and the total pressure of the system was less than 0.5 kbar. Using this relaxed structure, the *3N$_{atom}$* phonon frequencies were calculated at Γ. The three optical modes of BAs are both infrared (IR) and Raman-active. Though the material is slightly polar, the Born effective charges of B and As are small (~|0.59|), and cause weak longitudinal optical-transverse optical (LO-TO) splitting: 702.95 cm$^{-1}$ ($\omega_{TO}$) to 708.12 cm$^{-1}$ ($\omega_{LO}$). Our Born effective charges are in close agreement with Liu et al.[23] We subsequently used the Lyddane-Sachs-Teller relation to derive the static dielectric constant ($\varepsilon_0$) from the LO and TO phonon frequencies and the high-frequency dielectric constant ($\varepsilon_\infty$) as calculated within the random-phase approximation (RPA) using the GW method. Our results are listed in Table 3, along with the refractive index, $\sqrt{\varepsilon_\infty}$. We find that the two dielectric constants have very similar values, indicative of the small charge transfer between the B and As atoms. The similarity of the two values also indicates that the scattering of electrons by polar optical phonons[24] is weaker than other III-V semiconductors, which possibly leads to relatively higher electron and hole mobilities.[23]

Our calculated imaginary part of the dielectric function ($\varepsilon_2$) with and without excitonic effects for BAs is shown in Fig. 2. Since we only considered direct optical transitions, the absorption onset coincides with the minimum direct gap. The strong peak observed around 6 eV is due to the existence of parallel bands along the Γ-K, L-U, L-K, and U-X paths, which facilitate a higher joint density of states at this photon energy and lead to the spike in absorption. In previous work, Stukel also calculated the $\varepsilon_2$ spectrum, in good agreement with our results.[7] We find our main features are shifted to ~1 eV higher energies, corresponding to the gap opening due to our inclusion of GW corrections. The inclusion of the electron-hole interaction shifts the spectral weight to longer wavelengths not because of a redshift of transition energies, but due to the redistribution of the optical matrix elements, as previously reported in, e.g., Si.[25]

We also evaluate the exciton binding energy using the Bohr model, $E_b = 13.6$ eV $\mu^*/\varepsilon_\infty^2$, and our calculated material parameters. For the reduced mass $\mu^* = m_e m_h/(m_e + m_h)$ we used the directionally averaged values for the electron and heavy-hole masses. The resulting binding energy of 43 meV indicates that excitons are relatively stable against thermal dissociation at room temperature. The larger exciton binding compared to, e.g., bulk Si (15 meV)[26] is understood due to the heavier effective masses and smaller dielectric constant of BAs.

Table 3: Calculated material properties of zincblende BAs

| Property | Value |
|---|---|
| Indirect gap | 2.049 eV |
| Minimum direct gap | 4.135 eV |
| Ionization potential | 6.544 eV |
| Electron affinity | 4.495 eV |
| Spin-orbit splitting | 0.206 eV |
| $m_{e,l}^*/m_0$ | 1.093 ± 0.113 |
| $m_{e,t}^*/m_0$ | 0.239 ± 0.001 |
| $m_{soh,1}^*/m_0$ ($\Gamma \rightarrow K$) | 0.166 ± 0.003 |
| $m_{soh,2}^*/m_0$ ($\Gamma \rightarrow K$) | 0.191 ± 0.004 |
| $m_{lh,1}^*/m_0$ ($\Gamma \rightarrow K$) | 0.169 ± 0.003 |
| $m_{lh,2}^*/m_0$ ($\Gamma \rightarrow K$) | 0.220 ± 0.004 |
| $m_{hh,1}^*/m_0$ ($\Gamma \rightarrow K$) | 0.595 ± 0.059 |
| $m_{hh,2}^*/m_0$ ($\Gamma \rightarrow K$) | 0.982 ± 0.061 |
| $m_{soh}^*/m_0$ ($\Gamma \rightarrow L$) | 0.154 ± 0.003 |
| $m_{lh}^*/m_0$ ($\Gamma \rightarrow L$) | 0.136 ± 0.008 |
| $m_{hh}^*/m_0$ ($\Gamma \rightarrow L$) | 0.727 ± 0.003 |
| $m_{soh}^*/m_0$ ($\Gamma \rightarrow X$) | 0.258 ± 0.001 |
| $m_{lh}^*/m_0$ ($\Gamma \rightarrow X$) | 0.243 ± 0.001 |
| $m_{hh}^*/m_0$ ($\Gamma \rightarrow X$) | 0.253 ± 0.001 |
| $\varepsilon_0$ | 9.15 |
| $\varepsilon_\infty$ | 9.02 |
| Exciton Binding Energy | 43 meV |
| Refractive index | 3.00 |
| $\omega_{TO}$ | 702.95 cm$^{-1}$ |
| $\omega_{LO}$ | 708.12 cm$^{-1}$ |
| $a$ (experimental)[a] | 4.7776 Å |

[a] Value taken from Ref. 5, accessed via Inorganic Crystal Structure Database

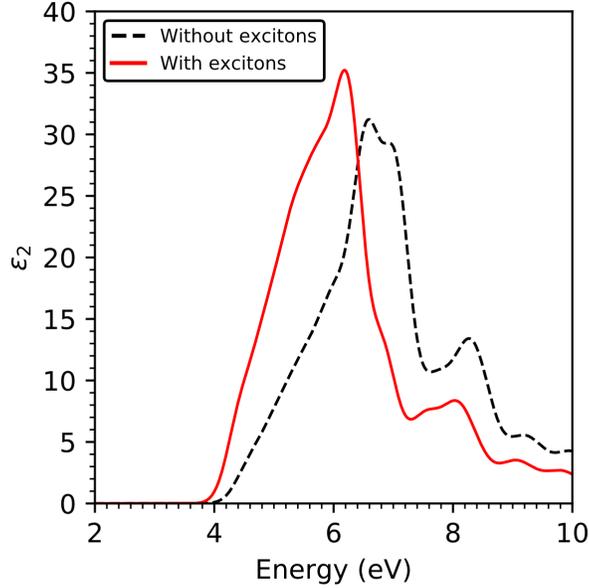

Figure 2: Imaginary dielectric function of BAs calculated within the GW method. The dashed black line and the solid red line correspond to $\varepsilon_2$ without and with excitonic effects, respectively.

In conclusion, we systematically characterized the electronic and optical properties of BAs with first-principles calculations that include quasiparticle and spin-orbit coupling corrections. We determined the fundamental band gap of BAs to be indirect with a value of 2.049 eV, while the minimum direct band gap is 4.135 eV. Our calculations show that spin-orbit effects split the topmost valence bands by 0.206 eV. Our effective-mass calculations find that holes are lighter than electrons in BAs, indicating that the mobility is higher for holes than electrons, and that acceptors are more easily ionized than donors. Our calculations of the Born effective charges and dielectric constants indicate that BAs is weakly ionic, a conclusion also reached by examining the orbital character of the VBM. Our results for the direct optical absorption spectrum show a strong peak for photon energies around 6 eV, which coincides with parallel valence and conduction bands and a large joint density of states at this particular photon energy. Our estimate for the exciton binding energy indicates that excitons are relatively stable at room temperature. Our work uncovers the fundamental electronic properties of this promising high-thermal-conductivity semiconductor to guide applications in electronic devices.

**Acknowledgements**
This work was supported by the Designing Materials to Revolutionize and Engineer our Future (DMREF) Program under Award No. 1534221, funded by the National Science Foundation. K.A.M. acknowledges the support from the National Science Foundation Graduate Research Fellowship Program through Grant No. DGE 1256260. This research used resources of the National Energy Research Scientific Computing Center, a DOE Office of Science User Facility supported by the Office of Science of the U.S. Department of Energy under Contract No. DE-AC02-05CH11231.